\documentclass[a4paper,fleqn,usenatbib]{mnras}

\usepackage[T1]{fontenc}
\usepackage{ae,aecompl}

\usepackage{graphicx}	
\usepackage{amsmath}	
\usepackage{amssymb}	

\def\th{\thinspace}

\title[A Detached Binary Star in the NGC 6791]{A Turnoff Detached Binary Star V568 Lyr in the {\it{Kepler}} Field of the Oldest Open Cluster (NGC 6791) in the Galaxy}

\author[K. Yakut et al.]{
K. Yakut,$^{1,2}$\thanks{E-mail: kadri.yakut@ege.edu.tr}
P. P. Eggleton,$^{1,2,3}$
B. Kalomeni,$^{1,4}$
C. A. Tout,$^{2}$
J. J. Eldridge$^{2,4}$
\\
$^{1}$Department of Astronomy and Space Sciences, University of Ege, 35100, Bornova-{\.I}zmir, Turkey\\
$^{2}$Institute of Astronomy, The Observatories, University of Cambridge, Madingley Road, Cambridge CB3 0HA, UK\\
$^{3}$Lawrence Livermore National Laboratory, 7000 East Ave, Livermore, CA 94551, USA\\
$^{4}$37-575, MIT Kavli Institute for Astrophysics and Space Research, 70 Vassar St, Cambridge, MA 02139, USA\\
$^{5}$Department of Physics, University of Auckland, Private Bag 92019, Auckland, New Zealand
}

\date{Accepted XXX. Received YYY; in original form ZZZ}

\pubyear{2015}

\begin{document}
\label{firstpage}
\pagerange{\pageref{firstpage}--\pageref{lastpage}}
\maketitle

\begin{abstract}
  We present the \emph{Kepler} photometric light-variation analysis of
  the late-type double-lined binary system V568~Lyr that is in the
  field of the high metallicity old open cluster NGC~6791.  The radial
  velocity and the high-quality short-cadence light curve of the
  system are analysed simultaneously.  The masses, radii and
  luminosities of the component stars are $M_1 = 1.0886\pm0.0031\,
  M{\odot}$, $M_2 = 0.8292 \pm 0.0026\, M{\odot}$, $R_1 = 1.4203\pm
  0.0058\, R{\odot}$, $R_2 = 0.7997 \pm 0.0015\, R{\odot}$, $L_1 =
  1.85\pm 0.15\, L{\odot}$, $L_2 = 0.292 \pm 0.018\, L{\odot}$ and
  their separation is $a = 31.060 \pm 0.002\, R{\odot}$.  The
  distance to NGC~6791 is determined to be $4.260\pm 0.290\,$kpc by
  analysis of this binary system.  We fit the components of this
  well-detached binary system with evolution models made with the
  Cambridge {\sc stars} and {\sc ev(twin)} codes to test low-mass
  binary star evolution.  We find a good fit with a metallicity of $Z = 0.04$
  and an age of $7.704\,$Gyr.  The standard tidal dissipation, included in
  {\sc ev(twin)} is insufficient to arrive at the observed circular
  orbit unless it formed rather circular to begin with.
\end{abstract}




\section{Introduction}

Accurate photometric and spectroscopic observations of binary systems
provide well-determined parameters of the components.  These
parameters, in turn, allow us to test stellar evolution models.  Light
variations of a large number of variable stars have been obtained by
successful observations of the \textit{Kepler} space telescope.  The
open stellar cluster NGC~6791 is one of the four open clusters that fall
in the observed field of the \textit{Kepler} project
\citep{2010ApJ...713L..79K}. It is the oldest open clusters in
the Galaxy and contains many close and wide binary systems. However,
only a few of them fall within {\it Kepler}'s view. Studies of binary
systems that are members of a cluster allow us to estimate both the
age of the host cluster and its distance.

Galactic clusters provide information about the evolution of the host
galaxy.  In this regard, old open clusters are crucial and NGC~6791,
Be~17, M67, NGC 188, etc.\
\citep{1997ApJ...483..826P,2008ApJ...678.1279B,2009AJ....137.5086M,KY09,2011ApJ...729L..10B,2011A&A...525A...2B}
are important old open clusters.  Binary systems that are member of
clusters with different chemical composition are good targets to test
how stellar evolution varies with composition, when we can determine
their physical parameters and construct appropriate evolutionary
models.  The system V568~Lyr resides at the main-sequence turn-off in
the colour--magnitude diagram (CMD) of the cluster and this turn-off point
gives us information on the age of the cluster. Hence, the study of
this particular system gives substantial information on both the stars
and the cluster in which they reside. This is a crucial system,
particularly given its detached configuration and accurate radial
velocity curve and photometry.

\citet{2009AJ....137.4949B} analysed the chemical composition of the
stars at the turn-off of NGC~6791 with Keck~+~HIRES spectra and
estimated the metallicity to be more than twice that of the Sun.
Recently, \citet{2014arXiv1412.8515B} repeated the study and similarly
estimated ${\rm [Fe/H]} = 0.30\pm 0.02$. Previous studies on the
metallicity of the cluster give the [Fe/H] ratio as $0.30-0.40$
\citep{1998ApJ...502L..39P,2006ApJ...646..499O}.
\citet{2014ApJ...796...68B} analysed archival Keck/HIRES spectra of main--sequence turn--off and evolved stars of NGC~6791. They obtained C, N, O and Na abundances.
There are various methods to determine the age of the cluster but determination of the
chemical composition is always an important factor.
Brogaard et al. (2011, 2012) used
binary system properties of V568 Lyr to estimate the age of the
cluster to be $8.3\pm0.3\,$Gyr consistent with other estimates of
$8-10\,$Gyr \citep{2012A&A...543A.106B,2014arXiv1412.8515B}.  There
have been a number of photometric studies of NGC~6791 and some of its
stars.  \citet{1994AJ....108..585M} observed the cluster with the
$UBVI$ filters of the $0.9\,$m KPNO telescope.  \citet{Stetson03}
obtained detailed $BVR$ observations with the $2.5\,$m Nordic Optical
Telescope and obtained data for colour--magnitude and colour--colour
diagrams. \citet{{2011ApJ...733L...1P}} present comprehensive cluster membership and
$g'r'$ photometry of the cluster. Using {\it Kepler} photometry, \citet{2012ApJ...757..190C} studied solar-like oscillations of red giants of the cluster.
Photometric studies of some of the variable stars in
NGC~6791 have also been made by \citet{2011A&A...525A...2B},
\citet{1993MNRAS.265...34K} and \citet{2007A&A...471..515D}.

V568~Lyr is a relatively long-period system so a full and accurate
light curve of the system could not be obtained.
\citet{1993MNRAS.265...34K} and \citet{1996MNRAS.282..705R} determined
its orbital period.  Later \citet{gru08} and
\citet{2011A&A...525A...2B} performed light and radial velocity
analyses of ground-based data and determined masses of
$1.0868\pm0.0039 \rm{M{\odot}}$ and $0.8276\pm0.0022 \rm{M{\odot}}$,
radii of $1.397\pm0.013 \rm{R{\odot}}$ and $0.7813\pm0.0053
\rm{R{\odot}}$ for the primary and secondary components.

The field of NGC~6791 is rich in binary systems, most of which are
members of the cluster.  The systems V519~Lyr, V522~Lyr and V564~Lyr
are interacting close binary systems. There is no published
spectroscopic study of these systems. V568~Lyr is, on the other hand,
a well-detached system and has excellent radial velocity curves
\citep{gru08,2011A&A...525A...2B}. This, together with the new
{\it{Kepler}} light curve, makes it important to determine accurate
physical parameters of its components and so provide better distance
and age estimates.  Here, owing to their high sensitivity and quality,
we use public {\it{Kepler}} observations supplemented with the radial
velocity and abundance measurements of \citet{gru08} and
\citet{2011A&A...525A...2B}. In the next section, we discuss the
public {\it{Kepler}} data and the light curve and radial velocity data
of V568~Lyr are solved simultaneously. In the third section, the
parameters of the system and the distance to the cluster are
determined.  In the fourth section, non-conservative binary
evolutionary models are constructed with {\sc ev(twin)} and then these
are compared with those of earlier studies.

\section{Analysis of the \textit{Kepler} Data and Light Curve Modelling}

\begin{table}
\caption{Basic properties of V568~Lyr (KIC~2437452).  Magnitudes, temperature and $E(B-V)$ are from
\citet{1994AJ....108..585M} and \citet{2011A&A...525A...2B}
and other parameters from the {\it Kepler} Input Catalogue (KIC) and SIMBAD.}
\begin{tabular}{llll}
\hline
Kepler ID                 &&& 2437452                   \\
2MASS ID                  &&& 19205427+3745347          \\
$\alpha_{2000}$           &&& 19:20:54.28               \\
$\delta_{2000}$           &&& 37:45:34.7                \\
2MASS \textit{J}                  &&& 15.513 mag            \\
2MASS \textit{H}                   &&& 15.057 mag            \\
2MASS \textit{Ks}                  &&& 15.076 mag           \\
\textit{K}$_p$(\textit{Kepler})    &&& 16.981 mag                \\
\textit{V}$_{12}$                  &&& 17.54 mag                \\
\textit{T}$_{\rm eff}$/K           &&& 5645             	      \\
\textit{E(B-V)}                    &&& 0.16                       \\
Spectral type                      &&& G5V + K3V             \\
\hline
\end{tabular}
\label{tab:data}
\end{table}

\begin{figure}
\includegraphics[width=\columnwidth]{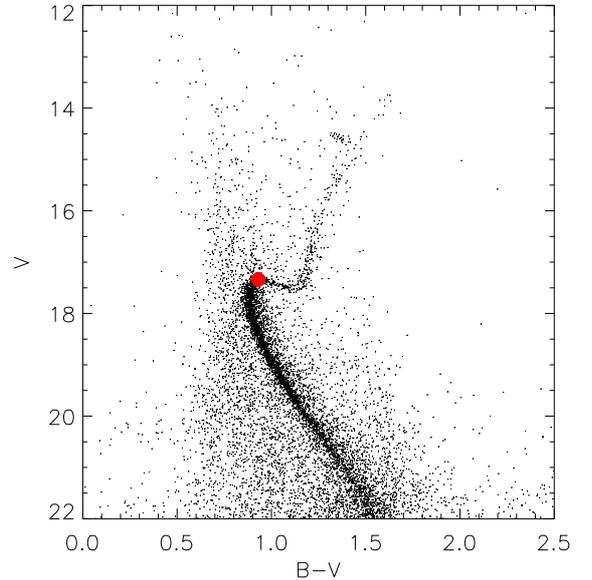}
\caption{The CMD of the open cluster NGC~6791.  Data are from \citet{Stetson03}.
The large filled circle indicates V568~Lyr, located at the
main-sequence turn-off in the cluster CMD.}\label{fig:ngc6791CMD}
\end{figure}

\begin{figure*}
\includegraphics[scale=0.6]{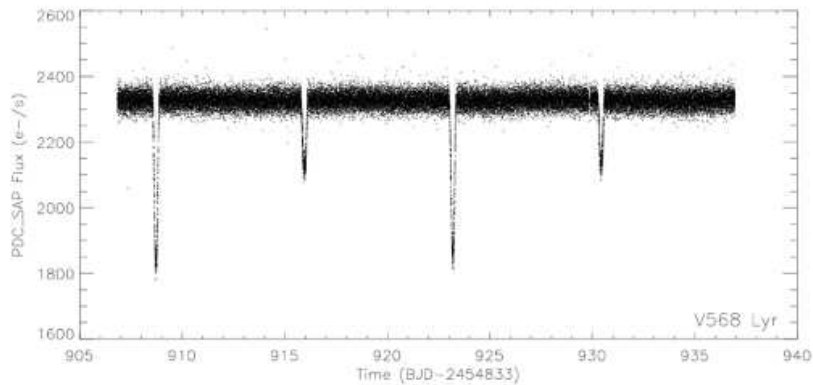}
\caption{Short-cadence \textit{Kepler} observations of V568~Lyr.}\label{fig:KeplerSC}
\end{figure*}

The \textit{Kepler} space telescope has observed nearly 150000 stars
with different properties, such as planetary components, pulsating
stars, etc. The accuracy of the \textit{Kepler} telescope is very
high.  Part of the NGC~6791 field is in the \textit{Kepler} field.
The binary system~V568 Lyr is shown on the CMD of NGC~6791 in Fig.~\ref{fig:ngc6791CMD}. Because it lies
right at the main-sequence turn-off, its analysis is very important
for the cluster as a whole. V568~Lyr was observed during quarters
$\rm Q1--Q17$ with long cadence (exposure time of about $30$\,min) and
one quarter (Q10) in short cadence ($1\,$min). The observed
parameters are summarized in Table~\ref{tab:data}. For the
light-curve analysis, we selected the short--cadence Q10 public data
that cover $30.11\,$d continuously. This covers
more than twice the orbital period. The short cadence \textit{Kepler} observations of V568 Lyr are shown in
Fig.~\ref{fig:KeplerSC}. The duration of the primary and secondary
minima are almost same and are about $7.8\,$h. The secondary minima
show a transit, and its duration is about $2.8\,$h. These properties
of the light curve allow us to determine rather accurate relative
radii of the components.

\begin{figure}
\includegraphics[width=\columnwidth]{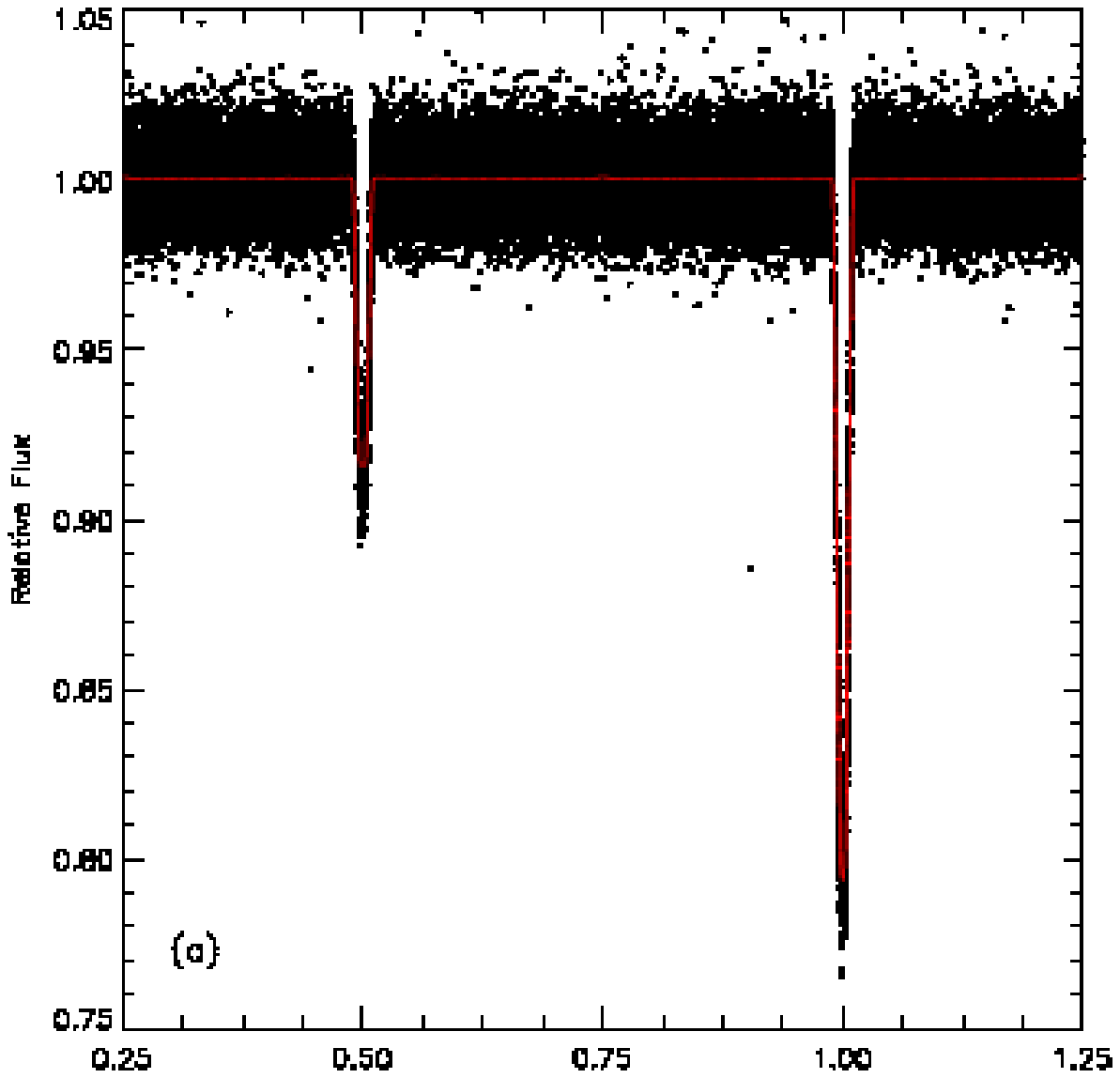}\\
\includegraphics[width=\columnwidth]{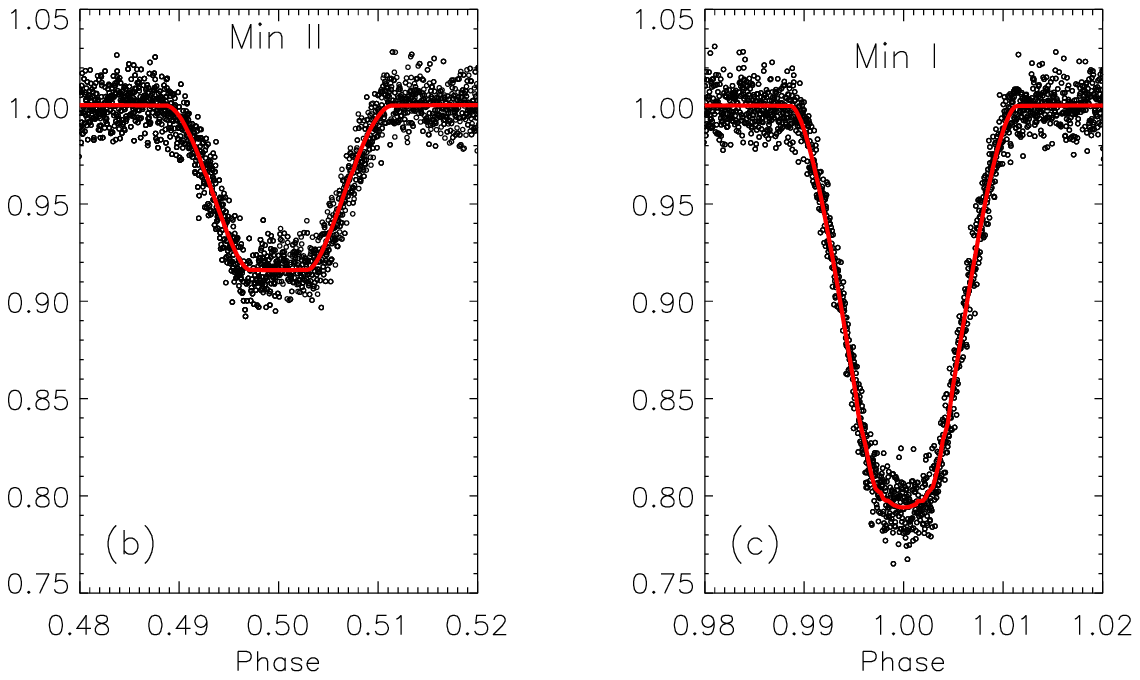}\\
\includegraphics[width=\columnwidth]{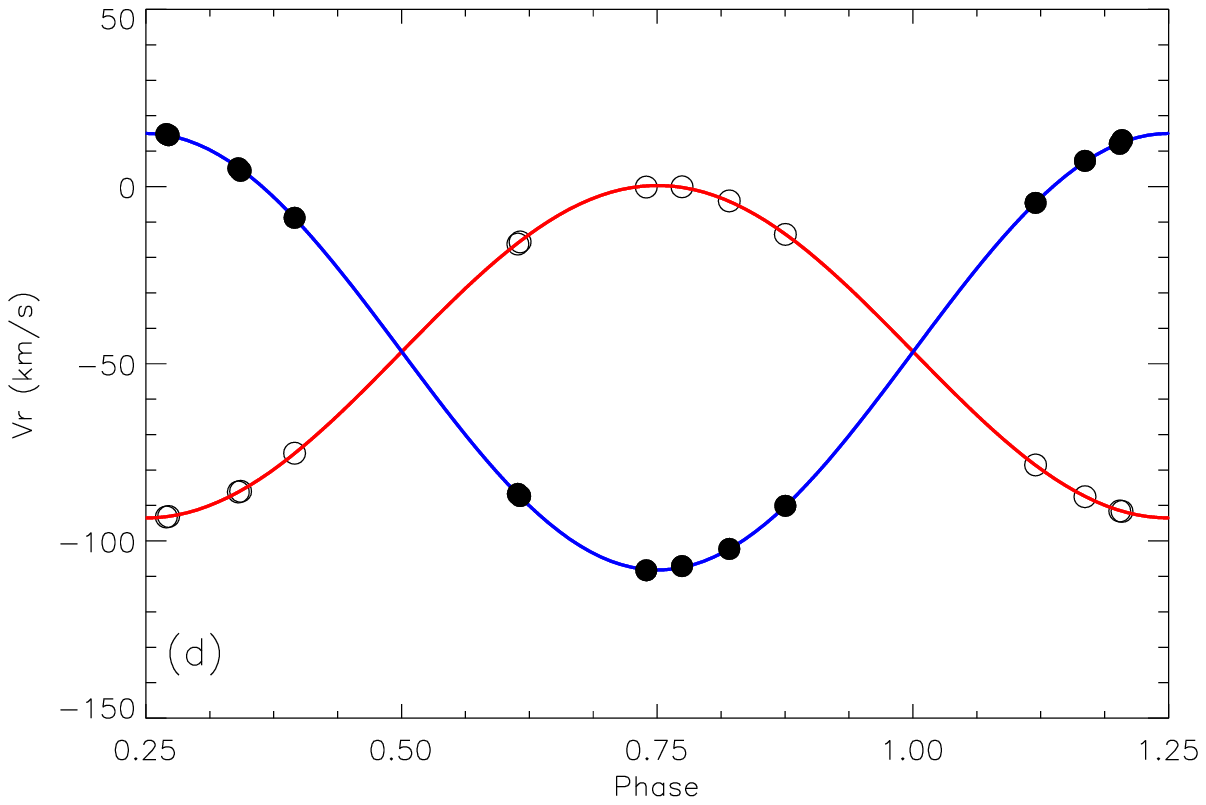}\\
\caption{(a) Observed (dots) and computed (line) light curve of V568 Lyr.
Primary (b) and secondary (c) minima expanded to emphasize the agreement.
(d) Observed and computed radial velocities.}\label{fig:V568Lyr}
\end{figure}

As well as the {\it Kepler} light curve, accurate radial velocity observations
of the system exist in the literature.  \cite{gru08} and \cite{2011A&A...525A...2B}
solved radial velocities with part of the light curve of V568~Lyr obtained with
ground-based telescopes. Here, we use the highly accurate and full light curve
obtained by {\it{Kepler}}. The system analysis
is made simultaneously with {\it{Kepler}} photometric observations and the
radial velocity curves. We expect to obtain more accurate orbital and physical
parameters.  The {\it Kepler} are solved
with the {\sc{phoebe}} code \citep{prs05}, based on the \citet{wilson71} code.
Initial parameters for the light--curve analysis are taken from \cite{gru08} and
\cite{2011A&A...525A...2B}. The gravity-darkening coefficients {$g_1$} and
{$g_2$} are obtained from \citet{lucy67}. Albedos {$A_1$} and
{$A_2$} are from \citet{ruc69} and the limb-darkening coefficients
from \citet{claret11}. The orbital inclination $i$, mass ratio
$q$, temperature of the secondary $T_2$, surface potentials
$\Omega_1$ and $\Omega_2$, luminosity $L_1$, phase shift $\phi$, the time
of minimum light $T_0$ and the orbital period $P$ are treated as
free parameters. Previous observations have suggested a third body in the system so,
during the light--curve analysis, the contribution of a third body to the total
light parameter $l_3$ is added as a free parameter.  With a detailed
solution, the orbital, geometric and radiative parameters we obtain are given in
Table~\ref{tab:lc}. The results obtained from observations (dots) and from our
light--curve model (solid line) are shown in Fig.~\ref{fig:V568Lyr}a.  To show the
accuracy of the solution the residuals and minima are expanded in
Figs~\ref{fig:V568Lyr}b-d.  In Fig.~\ref{fig:V568Lyr}e, the radial velocity curves
of the components and those calculated with simultaneous solutions
are shown.

\begin{table}
\caption{Photometric and spectroscopic elements of V568Lyr and their formal
1$\sigma$ errors.  See text for details.}\label{tab:lc}
\begin{tabular}{ll}
\hline
$T_{0}$ as ${\rm JD}/{\rm{d} - 2400000}$   & 55741.71981(46) \\
$P/{\rm d}$                               &14.47672(50)   \\
$i/^\circ$                                & 89.05(4)	  \\
$\Omega _{1}$                             & 21.71(5)	  \\
$\Omega _{2}$                             & 32.87(3)	  \\
$q = m_2/m_1$                             & 0.7616(1)    \\
$T_1/{\rm K}$                             & 5645          \\
$T_2/{\rm K}$                             & 4734(80)	  \\
Luminosity ratios:                        & \\
$l_1/l_\textrm{total}$                    & 74.5\%         \\
$l_2/l_\textrm{total}$                    & 8.5\%      \\
$l_3/l_\textrm{total}$                    & 17.0\%    \\
Fractional radius of primary ($R_1/a$)    & 0.04568(11)   \\
Fractional radius of secondary ($R_2/a$)  & 0.02572(10)   \\
$K_1/{\rm km\,s^{-1}}$                   & 46.92(9)       \\
$K_2/{\rm km\,s^{-1}}$                    & 61.60(12)      \\
System velocity $V_\gamma/{\rm km\,s^{-1}}$  &$-46.6(1)$       \\
$e$                                       & 0.000(0.001)      \\
$a/R{\odot}$                             & 31.060(2)     \\
\hline
\end{tabular}
\end{table}

\section{Physical parameters of the system}
Light curve analysis of V568~Lyr gives radii, as fractions of the separation,
$R_{1}/a = 0.04568$ and $R_2/a = 0.02572$.  These small ratios mean that the components
closely preserve their spherical symmetry.  Because of the high sensitivity of
the {\it{Kepler}} data, the light curve of the system gives us the opportunity
to determine much more accurately its inclination and especially the radii of
the components relative to previous studies. Third light is detected
at the level of 17\,per cent.  The reddening is computed, from $E(B-V)$ in Table 1, to be
$A_V = 0.49$.  Radial velocity curves yield the velocity amplitudes of the
components as listed in Table~\ref{tab:lc}.

\begin{table}
\begin{center}
\caption{Absolute parameters of V568~Lyr.  Standard errors of 1$\sigma$ in the last digits are given in
parentheses.}\label{tab:V568Lyr:PhyPar}
\begin{tabular}{llll}
\hline
                                               & Primary           & Secondary   \\
\hline
Masses $M/\rm{M{\odot}}$                        & $1.0886(31)$      & $0.8292(26)$      \\
Radii $R/\rm{R{\odot}}$                         & $1.4203(58)$      & $0.7997(15)$      \\
Effective temperatures $T_{\rm eff}/{\rm K}$    & $5\,645(95)$      & $4\,734(80)$    \\
Luminosities $L/{\rm L{\odot}}$                 & $1.846(15)$       & $0.292(18)$      \\
Surface gravity $\log_{10}(g/\rm{cm\,s^{-2}})$   & 4.170(44)         & 4.551(33)       \\
Absolute bolometric magnitude $M_B$           & 4.09(7) mag         & 6.09(7) mag           \\
Absolute visual magnitude $M_V$               & 4.19(8) mag         & 6.53(8) mag      \\
Separation between stars $a/\rm{R_{\odot}}$      &~~~~~~~~~~~~31.06(3) &      \\
Distance $d/{\rm pc}$                          &~~~~~~~~~~~~4260(290) &      \\
\hline
\end{tabular}
\end{center}
\end{table}

When we leave the eccentricity as a free parameter during the
solution it tends to zero. Using the orbital parameters (Table~\ref{tab:lc})
we obtained the physical parameters of V568 Lyr and list them in
Table~\ref{tab:V568Lyr:PhyPar}. While solving, the temperature of the Sun is
taken to be 5772~K and its bolometric magnitude to be 4.755 mag (\citet{mamajek15}).
The temperatures of the components with the \citet{1996ApJ...469..355F}
tables give the bolometric corrections as BC$_1=-0.10$ mag, and BC$_2 = -0.44$ mag.
Using these BC values for \textit{V} band, we then obtain the absolute magnitudes of the components.

These new estimates of the system parameters, with the magnitudes and reddening, reveal the
distance of the system to be $4.26\pm0.29\,$kpc.  The masses we find
are consistent with those found by \citet{2011A&A...525A...2B}.
The radius of the primary star, on the other hand, is estimated to be $7\,$per
cent larger while the secondary star is estimated $4\,$per cent smaller than
in their study.  These differences are most probably due to the higher accuracy
of {\it{Kepler}} data relative to ground-based observations.

\section{MODELLING and DISCUSSION}

We have simultaneously solved the radial velocity and light curve of the detached and relatively long
period binary V568~Lyr and determined its orbital and physical
parameters (Table~\ref{tab:V568Lyr:PhyPar}). The well-detached
configuration of V568~Lyr allows us to determine very accurate orbital and
physical parameters (Tables~\ref{tab:lc} and~\ref{tab:V568Lyr:PhyPar}).
The distance to the cluster is found to be 4.26$\pm 0.29$\,kpc, equivalent to a
distance modulus of $13.14\pm0.22$ mag. This is roughly
consistent with other distance estimates.  \citet{Harris81} found a distance
modulus of $14.0\pm0.2$ by analysing a CMD.
\citet{1985ApJ...291..595A} found 12.6 using isochrones.
\citet{2006ApJ...643.1151C} obtained 12.6\,--\,13.6 again with isochrones.
Based on the fit to the white dwarf cooling sequence \citet{2008ApJ...679L..29B} obtained distance scale as 13.50.
\citet{2011A&A...525A...2B} determined $13.51\pm0.06$ with binary system
parameters.  \citet{2011ApJ...729L..10B} measured $13.11\pm0.06$ for the red
giant stars.

To model this system and compare it with observations allows us to test our
understanding of stellar evolution.  V568~Lyr, in particular, gives us the opportunity to
estimate the age of the cluster.  We made with the {\sc ev} code
\citep{EKE02} and its much more powerful {\sc twin} variant \citep{YE05,ppe06,ppe10},
both of which are based on the Cambridge {\sc stars} code \citep{ppe71,ppe72,ppe73,Pols95}.

In single-star evolution, the effects of rotation and dynamo activity
on mass--loss and the proximity of the companion are often not
considered. The {\sc ev} code admits various non-conservative
processes to be applied to the primary component of the binary system.
In the newer {\sc twin} variant, {\it both} components are solved {\it
  simultaneously} so that the effect of tidal dissipation on stellar
rotation, orbital period and eccentricity and on dynamo activity, and
hence on mass--loss, in both components are treated self-consistently.
Stellar winds carry off angular momentum by way of magnetic braking
and this angular momentum can be extracted from the {\it orbit} if
tidal dissipation is strong enough. We might imagine that tidal
dissipation has played a role in the evolution of V568~Lyr because its
orbit is circular and many systems with $P > 5\,$d are markedly
eccentric.  Tidal dissipation leads to spin--orbit synchronization (or
pseudo-synchronization in eccentric orbits) much more quickly than it
causes circularization.

It is not easy to determine the initial state of a binary system if
the evolution is not conservative.  We can reasonably assume that the
initial masses were larger than now but it is not clear, without some
trial and error, by how much they were larger. Further, it is
entirely likely that there is some non-linearity in the behaviour so
that a straight-forward convergence procedure that assumes
near-linearity may not in fact converge or at least converge badly.

The metallicity of NGC 6791 has been estimated to be about two and a half times solar
\citep{gru08}, rather surprisingly because the cluster is also estimated by them to be
nearly twice the age of the Sun. However the metallicity of the Sun has
itself been subject to recent downward revision \citep{Asplund05} although there is
probably still some uncertainty because the new value seems to spoil the apparent agreement
of earlier models of the Sun with helioseismology \citep{Bahcall05}. We have used the
tabular data of the OPAL data base \citep{Iglesias96} for $Z=0.04$,
corresponding to nearly three times the value that \citet{Asplund05} obtained for the Sun,
and to about 2.5 times the solar metallicity as estimated by more recently by Caffau
et al. (2011). However we adopt this value mainly because it leads to much better
agreement between the observational and theoretical luminosities, radii and masses
of the components, as illustrated in Fig. 4. The zero-age helium abundance was taken
to be $Y=0.32$, and the mixing-length ratio to be 2.0. Aspects of the code have been
described in detail by Eggleton (2006) and papers referred to therein. The only
(slight) novelty is that convective core overshooting has been calibrated fairly
carefully in a study by Eggleton \& Griffin (2015) of 53 binaries almost
all of which are evolved to giant or supergiant dimensions, where the issue of
convective overshooting is much more challenged than by near-main-sequence stars
including the present subject. The new model of overshooting in fact makes very
little difference in the present case.

After some experimentation, we evolved a pair of stars with initial masses $1.0994$ and $0.8603\,M{\odot}$,
spin periods of 2 d each, an eccentricity of 0.3 and an orbital period of 15.79 days.
We expect the orbital period to decrease if the orbit circularizes, but it is also
influenced by magnetic braking and by mass loss. In addition, it will also increase as
the orbit acquires  angular momentum from the spins of the stars by tidal friction. The
particular model of the non-conservative processes in the code (\citet{EKE02}, but see below)
led the system to evolve, after 7.733 Gyr to roughly the observed period, masses, radii
and temperatures
(cf Table~\ref{tab:V568Lyr:PhyPar}): 14.42d, 1.0888 and 0.8285\,$M{\odot}$, 1.428 and 0.781\,
$R{\odot}$,  1.949 and 0.309\,$L{\odot}$, and temperatures 5712 and 4872\,K.
We see that the secondary component model is rather too hot, large and luminous, but probably
not unacceptably so (2 -- 3\%). Fig. 4 shows the mean observed values ($\log L$ in solar
units and effective $\log T$ in kelvin) as black squares,
with an `error cloud' of dots distributed according to an assumed Gaussian distribution
with the standard errors as tabulated above. The blue circle is a point on the evolutionary
track of the primary that is reasonably close to the observed value, and the blue
asterisk is the co\ae val point on the secondary's (very short) track, i.e. a point on the
same isochrone as the primary.
\begin{figure}
\hskip -0.2truein\includegraphics[scale=0.65]{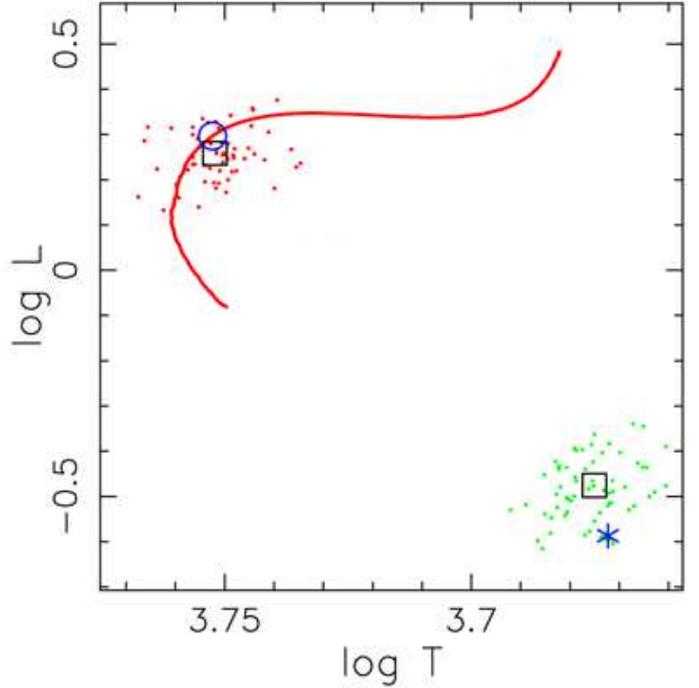}\\
\vskip -0.2truein
\caption{HRD of an evolutionary model with the TWIN code. Red, primary;
green secondary. Black squares: mean observed values. Dots: scatter due to
random measurement errors; see text. Blue circle: a well-fitting model value of the primary. Blue
asterisk: co-eval point for secondary.} \label{fig:TWINplot}
\vskip -0.2truein
\end{figure}
However, in order to produce a nearly circular orbit we had to increase the model tidal
friction by a factor of 100, and even then the eccentricity, starting from a hypothetical
0.3, was only reduced by a factor of 6, which might just be compatible with the observations.
We believe that the inadequacy of tidal friction is likely to be a problem
for any other model, because 14.5 d is a rather long period for two near-main-sequence stars
to have circularized their orbit, even in quite an old cluster \citep{ppe06}.

The fact that the two masses decreased by as much as $1$ per cent and $3$ per cent in the course of
$8\,$Gyr is perhaps noteworthy. Single stars similar to either component would lose
less mass but that is because they would have spun down to much slower rotation rates
by magnetic braking in the absence of tidal friction. The same non-conservative model as used
above, when applied to the Sun, gives the observed solar-wind strength and Alfv{\'e}n radius,
by design, at age $4.57\th$Gyr. This model predicts a loss of $1.4$ per cent of the initial mass
but almost all of this takes place in the first Gyr when the Sun is rotating in about
$14\th$d or less. The lower-mass component of our binary loses mass faster, partly because its
deeper convective layer is assumed to make it more active and partly because old K dwarfs are
normally rotating even more slowly than the Sun. This one is rotating substantially faster
if it is synchronized, which it surely must be.

\section*{Acknowledgements}

We acknowledge the {\it{Kepler}} Team for giving public access to their corrected light curves and thank the
anonymous referee for the careful reading and for suggestions. This research has made use of the SIMBAD
database, operated at CDS, Strasbourg, France.
This study was supported by the Turkish Scientific and Research Council (T\"UB\.ITAK 111T270, 112T766 and 113F097).
CAT thanks Churchill College for his fellowship.





\begin{thebibliography}{99}
\bibitem[\protect\citeauthoryear{Anthony-Twarog\& Twarog}{1985}]{1985ApJ...291..595A} Anthony-Twarog B.~J., Twarog B.~A., 1985, ApJ, 291, 595
\bibitem[\protect\citeauthoryear{Asplund, Grevesse, \& Sauval}{2005}]{Asplund05} Asplund M., Grevesse N., Sauval A.~J., 2005, ASPC, 336, 25
\bibitem[\protect\citeauthoryear{Bahcall, Serenelli, \& Basu}{2005}]{Bahcall05} Bahcall J.~N., Serenelli A.~M., Basu S., 2005, ApJ, 621, L85
\bibitem[\protect\citeauthoryear{Basu et al.}{2011}]{2011ApJ...729L..10B}Basu S., et al., 2011, ApJ, 729, LL10
\bibitem[\protect\citeauthoryear{Bedin et al.}{2008b}]{2008ApJ...679L..29B} Bedin L.~R., Salaris M., Piotto G., Cassisi S., Milone A.~P., Anderson J., King I.~R., 2008b, ApJ, 679, L29
\bibitem[\protect\citeauthoryear{Bedin et al.}{2008a}]{2008ApJ...678.1279B} Bedin L.~R., King I.~R., Anderson J., Piotto G., Salaris M., Cassisi S., Serenelli A., 2008a, ApJ, 678, 1279
\bibitem[\protect\citeauthoryear{Bragaglia et al.}{2014}]{2014ApJ...796...68B} Bragaglia A., Sneden C., Carretta E., Gratton R.~G., Lucatello S., Bernath P.~F., Brooke J.~S.~A., Ram R.~S.,
2014, ApJ, 796, 68
\bibitem[\protect\citeauthoryear{Boesgaard, Jensen, \& Deliyannis}{2009}]{2009AJ....137.4949B} Boesgaard A.~M., Jensen E.~E.~C., Deliyannis C.~P., 2009, AJ, 137, 4949
\bibitem[\protect\citeauthoryear{Boesgaard, Lum, \& Deliyannis}{2014}]{2014arXiv1412.8515B} Boesgaard A.~M., Lum M.~G., Deliyannis C.~P., 2014, arXiv, arXiv:1412.8515
\bibitem[\protect\citeauthoryear{Brogaard et al.}{2011}]{2011A&A...525A...2B} Brogaard K., Bruntt H., Grundahl F., Clausen J.~V., Frandsen S., Vandenberg D.~A., Bedin L.~R., 2011, A\&A, 525, A2
\bibitem[\protect\citeauthoryear{Brogaard et al.}{2012}]{2012A&A...543A.106B} Brogaard K., et al., 2012, A\&A, 543, A106
\bibitem[\protect\citeauthoryear{Caffau et al.}{2011}]{2011SoPh..268..255C} Caffau E., Ludwig H.-G., Steffen M., Freytag B., Bonifacio P., 2011, SoPh, 268, 255
\bibitem[\protect\citeauthoryear{Carraro etal.}{2006}]{2006ApJ...643.1151C} Carraro G., Villanova S., Demarque P., McSwain M.~V., Piotto G., Bedin L.~R., 2006, ApJ, 643, 1151
\bibitem[\protect\citeauthoryear{Claret \& Bloemen}{2011}]{claret11} Claret, A., Bloemen, S., 2011, A\&A, 529, 75
\bibitem[\protect\citeauthoryear{Corsaro et al.}{2012}]{2012ApJ...757..190C} Corsaro E., et al., 2012, ApJ, 757, 190
\bibitem[\protect\citeauthoryear{de Marchi et al.}{2007}]{2007A&A...471..515D} de Marchi F., et al., 2007, A\&A, 471, 515
\bibitem[Eggleton \& Kiseleva-Eggleton (2002)]{EKE02} Eggleton, P.~P.,  \& Kiseleva-Eggleton, L.\ 2002, ApJ, 575, 461
\bibitem[Eggleton (1971)]{ppe71} Eggleton P.~P., 1971, MNRAS, 151, 351
\bibitem[Eggleton (1972)]{ppe72} Eggleton P.~P., 1972, MNRAS, 156, 361
\bibitem[Eggleton (1973)]{ppe73}Eggleton P.~P., 1973, MNRAS, 163, 279
\bibitem[\protect\citeauthoryear{Eggleton}{2006}]{ppe06} Eggleton P., 2006, Evolutionary Processes in Binary and Multiple Stars, Cambridge Univ. Press, Cambridge
\bibitem[\protect\citeauthoryear{Eggleton}{2010}]{ppe10} Eggleton P.~P., 2010, New Astron. Rev., 54, 45
\bibitem[\protect\citeauthoryear{EggletonGriffin}{2015}]{ppegrif15} Eggleton P.~P. \& Griffin, R.~E. 2015, in preparation
\bibitem[\protect\citeauthoryear{Flower}{1996}]{1996ApJ...469..355F} Flower P.~J., 1996, ApJ, 469, 355
\bibitem[\protect\citeauthoryear{Grundahl et al.}{2008}]{gru08} Grundahl F., Clausen J.~V., Hardis S., Frandsen S., 2008, A\&A, 492, 171
\bibitem[\protect\citeauthoryear{Harris \& Canterna}{1981}]{Harris81} Harris W.~E.,Canterna R., 1981, AJ, 86, 1332
\bibitem[\protect\citeauthoryear{Rogers, Swenson, \& Iglesias}{1996}]{Iglesias96} Rogers F.~J., Swenson F.~J., Iglesias C.~A., 1996, ApJ, 456, 902
\bibitem[\protect\citeauthoryear{Kaluzny \& Rucinski}{1993}]{1993MNRAS.265...34K} Kaluzny J., Rucinski S.~M., 1993, MNRAS, 265, 34
\bibitem[\protect\citeauthoryear{Koch et al.}{2010}]{2010ApJ...713L..79K} Koch D.~G., et al., 2010, ApJ, 713, L79
\bibitem[\protect\citeauthoryear{Lucy}{1967}]{lucy67} Lucy, L. B. 1967, Zeit. Astrophys., 65, 89
\bibitem[\protect\citeauthoryear{Pecaut \& Mamajek}{2013}]{mamajek15} Pecaut, M.~J. \& Mamajek E.~E., 2013,ApJS, 208, 9
\bibitem[\protect\citeauthoryear{Meibom et al.}{2009}]{2009AJ....137.5086M} Meibom S., et al., 2009, AJ, 137, 5086
\bibitem[Montgomery et al.(1994)]{1994AJ....108..585M} Montgomery, K.~A., Janes, K.~A., \& Phelps, R.~L.\ 1994, AJ, 108, 585
\bibitem[\protect\citeauthoryear{Origlia et al.}{2006}]{2006ApJ...646..499O} Origlia L., Valenti E., Rich R.~M., Ferraro F.~R., 2006, ApJ, 646, 499
\bibitem[\protect\citeauthoryear{Peterson \& Green}{1998}]{1998ApJ...502L..39P} Peterson R.~C., Green E.~M., 1998, ApJ, 502, L39
\bibitem[\protect\citeauthoryear{Phelps}{1997}]{1997ApJ...483..826P} Phelps R.~L., 1997, ApJ, 483, 826
\bibitem[\protect\citeauthoryear{Rucinski}{1969}]{ruc69} Rucinski, S. M. 1969, Acta Astron. 19, 245
\bibitem[\protect\citeauthoryear{Rucinski, Kaluzny, \& Hilditch}{1996}]{1996MNRAS.282..705R} Rucinski S.~M., Kaluzny J., Hilditch R.~W., 1996, MNRAS, 282, 705
\bibitem[\protect\citeauthoryear{Platais et al.}{2011}]{2011ApJ...733L...1P} Platais I., Cudworth K.~M., Kozhurina-Platais V., McLaughlin D.~E., Meibom S., Veillet C., 2011, ApJ,
733, L1
\bibitem[\protect\citeauthoryear{Pols et al.}{1995}]{Pols95} Pols O.~R., Tout C.~A., Eggleton P.~P., Han Z., 1995, MNRAS, 274, 964
\bibitem[\protect\citeauthoryear{Pr{\v{s}}a \& Zwitter}{2005}]{prs05} Pr{\v{s}}a, A., Zwitter, T., 2005, ApJ, 628, 426
\bibitem[\protect\citeauthoryear{Stetson, Bruntt, \& Grundahl}{2003}]{Stetson03} Stetson P.~B., Bruntt H., Grundahl F., 2003, PASP, 115, 413
\bibitem[\protect\citeauthoryear{Wilson \& Devinney}{1971}]{wilson71} Wilson, R.E., Devinney, E.J., 1971, ApJ, 166, 605
\bibitem[\protect\citeauthoryear{Yakut \& Eggleton}{2005}]{YE05} Yakut K., Eggleton P.~P., 2005, ApJ, 629, 1055
\bibitem[\protect\citeauthoryear{Yakut et al.}{2009}]{KY09} Yakut K., et al., 2009, A\&A, 503, 165



\end{thebibliography}



\bsp	
\label{lastpage}
\end{document}